\RequirePackage[2020-02-02]{latexrelease}

\documentclass[twocolumn,preprintnumbers]{revtex4}%
\usepackage{amsopn}
\usepackage{amssymb}
\usepackage{amsmath}
\usepackage{units}
\usepackage{graphicx}
\usepackage{epstopdf}
\usepackage{dcolumn}
\usepackage{bm}
\usepackage{xcolor}
\usepackage{ifthen}
\usepackage[normalem]{ulem}
\usepackage{amsfonts}%
\RequirePackage[2020-02-02]{latexrelease}
\UseRawInputEncoding
\providecommand{\U}[1]{\protect\rule{.1in}{.1in}}
\providecommand{\U}[1]{\protect\rule{.1in}{.1in}}
\def\showal{1}
\newcommand{\al}[1]{\ifthenelse{\showal=1}{\textcolor{orange}{[[#1]]}}{}}

\newcommand{\eb}[1]{\ifthenelse{\showal=1}{\textcolor{blue}{Comment to be removed: #1}}{}}

\begin{document}
\title{A fiber-loop laser with a cooled erbium-doped section}
\author{Eyal Buks}
\email{eyal@ee.technion.ac.il}
\affiliation{Andrew and Erna Viterbi Department of Electrical Engineering, Technion, Haifa
32000, Israel}
\author{Boris A. Malomed}
\email{malomed@tauex.tau.ac.il}
\affiliation{Instituto de Alta Investigaci\'{o}n, Universidad de Tarapac\'{a}, Casilla 7D,
Arica, Chile}
\date{\today }

\begin{abstract}
An optical comb featuring an unusual structure has
been recently observed in a fiber-loop laser running at cryogenic
temperatures. In the present work, the effect of optical polarization on
the operation of the laser is explored. Multi--mode
lasing with a relatively high degree of inter--mode coherency is
experimentally demonstrated. The setup can be used for applications to
spectroscopy, communications, and quantum data storage.

\end{abstract}
\maketitle

\textbf{Introduction} -- Erbium doped fibers (EDFs) are widely employed in
many applications. An important aspect of the work with EDF is that its key
properties may strongly depend on the temperature
\cite{Desurvire_246,Nakazawa_613,Thevenaz_22,Kobyakov_1,Le_3611,Aubry_2100002,Xi_1}%
. In particular, at cryogenic temperatures, an EDF can be used for storing
quantum information
\cite{Wei_2209_00802,Ortu_035024,Liu_2201_03692,Bornadel_2412_16013,Tittel_2501_06110,Jing_031304,Gupta_044029}%
. Storage times exceeding 10 seconds \cite{Saglamyurek_241111,Shafiei_F2A}
have been achieved by implementing the method of persistent spectral hole
burning. Furthermore, a properly cooled EDF can support robust multi--mode
lasing regime. The latter feature is important as multimode lasing finds
diverse applications to spectroscopy, signal processing, and communications
\cite{Haken_laser,Semenov_23}. Multimode lasing in the telecom band has been
demonstrated by an EDF cooled with liquid nitrogen
\cite{Yamashita_1298,Liu_102988,Lopez_085401}. At still lower temperatures, an
unequally spaced optical comb (USOC) can be generated by a fiber loop laser
with an integrated EDF \cite{Buks_L051001}, see Fig. \ref{FigID}(b) below.
However, the mechanism responsible for the formation of the USOC in that setup
remains unknown. Actually, the USOC may be considered as the extreme opposite
to a regular optical comb, in which the frequency spacings between adjacent
peaks are all identical. In contrast, the frequency spacings between all pairs
of USOC peaks are unique.

\begin{figure}[ptb]
\includegraphics[width=3.2in,keepaspectratio]{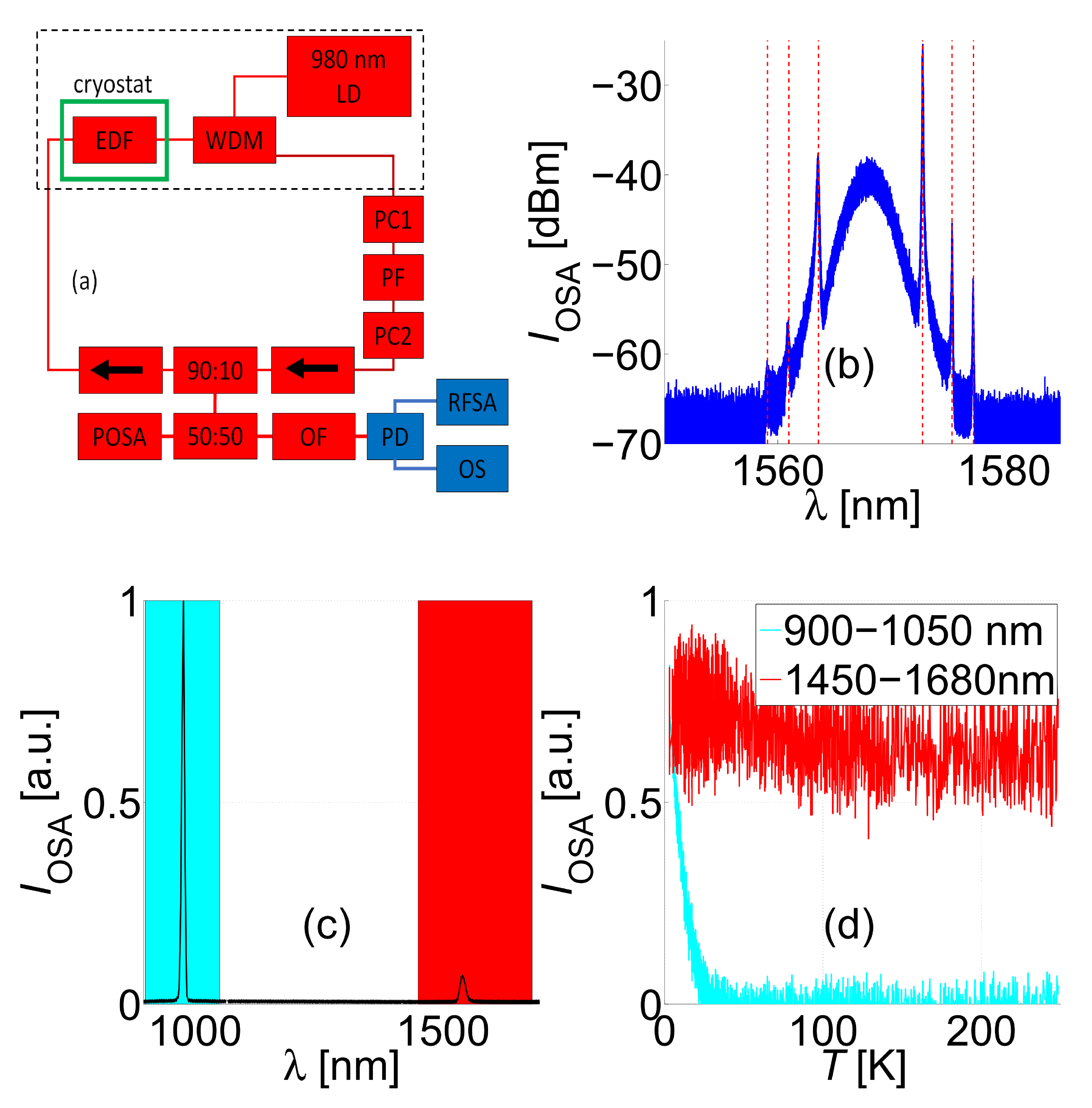} \caption{{} The
experimental setup. (a) Optical components and fibers are red colored, whereas
the blue color is used to label the RF components and coaxial cables. The
dashed rectangle labels the part of the setup used for the open-loop
transmittance measurement, which are presented in panels (c) and (d). (b)
POSA measurement of the optical spectrum with diode
current of $I_{\mathrm{D}}=0.8 \operatorname{A}$ (the
corresponding diode power is $P_{\mathrm{D}}=480 \operatorname{mW}$). The Kelly sidebands [see Eq. (\ref{Kelly sideband})] are labeled by overlaid vertical dashed
lines. (c) Optical spectrum at temperature $2.9 \operatorname{K}$. This
measurement is performed in the open-loop configuration, with a low-resolution
OSA connected to the left end of the EDF. The diode current is $I_{\mathrm{D}%
}=0.25 \operatorname{A}$ and the power is $P_{\mathrm{D}}=160
\operatorname{mW}$. (d) The integrated optical spectrum as a
function of temperature for the pumping band $900-1050 \operatorname{nm}$ [see
the cyan rectangle in (c)] and telecom band $1450-1680 \operatorname{nm}$ [see
the red rectangle in (c)].}%
\label{FigSetup}%
\end{figure}

The current study addresses effects of optical
polarization on the operation of the laser. A high
degree of polarization (DOP) is propitious for some potential USOC--based
applications. We explore the effect of polarization by integrating a
polarization filter (PF) and polarization controllers (PC) into the fiber
loop, see Fig. \ref{FigSetup} (a). We find that the inserted components enable
USOC formation with a relatively high DOP, as shown in Fig. \ref{FigDOP}. The
enhanced DOP is used for the generation of frequency-tunable radio-frequency
(RF) tones through coherent beating between pairs of USOC peaks, as shown in
Fig. \ref{FigRFSAOA} below. Moreover, the inserted polarization-controlling
ingredients allow tuning the fiber loop into a mode--locking phase
\cite{Hofer_720,Fermann_894}, see Fig. \ref{FigOffP} below.

\textbf{Experimental setup} -- A sketch of the unidirectional fiber loop laser
under the study is presented in Fig. \ref{FigSetup}(a). The loop is made of
$l_{\mathrm{EDF}}=5\operatorname{m}$ long EDF, and an undoped single-mode
fiber (Corning 28). The EDF absorption is $30$ dB $^{-1}$ at
$1530\operatorname{nm}$, and the mode field diameter is $6.5\mu
\operatorname{m}$ at $1550\operatorname{nm}$. A wavelength-division
multiplexing (WDM) device is used to integrate the EDF, which is installed
inside a cryostat, with a unidirectional room-temperature fiber loop. The EDF
is pumped by a $980\operatorname{nm}$ laser diode (LD), biased with current
$I_{\mathrm{D}}$. A 10:90 optical coupler (OC), and two isolators [labeled by
arrows in the sketch shown in Fig. \ref{FigSetup}(a)], are integrated into the
fiber loop. The output port of the 10:90 OC is connected to a polarimeter
optical spectrum analyzer (POSA) \cite{Buks_486}, and to a photo-detector
(PD), which is probed by both a radio-frequency spectrum analyzer (RFSA) and
an oscilloscope (OS). An optional optical filter (OF) with a central
wavelength $\lambda_{\mathrm{F}}$ and linewidth $1.2\operatorname{nm}$ (the
full width at half maximum) can be installed between the 10:90 OC and the PD.
The OF is installed and used only for the measurements presented below in Fig.
\ref{FigRFSAOA}.

The dependence on diode current $I_{\mathrm{D}}$ is shown in Fig. \ref{FigID}.
For all measurements presented by Fig. \ref{FigID}, the EDF temperature is
$2.8\operatorname{K}$. The measured PD averaged voltage $V_{\mathrm{PD}}$
(proportional to the averaged optical power) is plotted in Fig. \ref{FigID}(a)
as a function of $I_{\mathrm{D}}$. Below a critical
temperature of about $14\operatorname{K}$ USOC becomes visible in the optical
spectrum. The USOC sequence shown in Fig. \ref{FigID}(b) is measured with
diode current of $I_{\mathrm{D}1}$ [see Fig. \ref{FigID}(a)].

\begin{figure}[ptb]
\begin{center}
\includegraphics[width=3.2in,keepaspectratio]{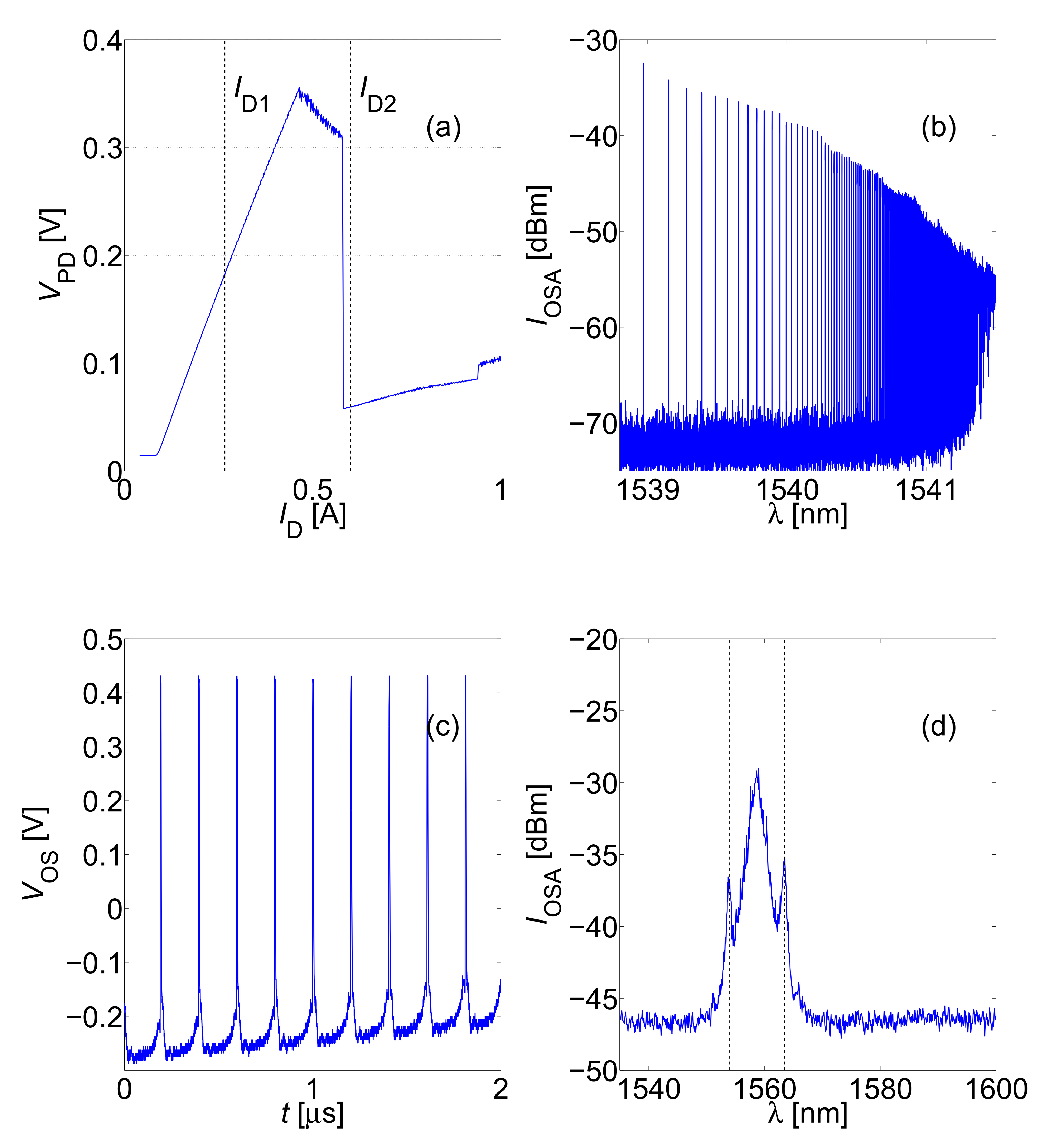}
\end{center}
\caption{{}Fiber-loop. The temperature is $2.8\operatorname{K}$. (a) The
dependence of the measured PD averaged voltage $V_{\mathrm{PD}}$ on diode
current $I_{\mathrm{D}}$. (b) USOC optical spectrum. (c) Optical pulses
measured by the OS. (d) Optical spectrum in the mode--locking regime. The
diode current is $I_{\mathrm{D}1}$ in (b) and $I_{\mathrm{D}2}$ in (c) and
(d).}%
\label{FigID}%
\end{figure}

\textbf{USOC} -- The underlying mechanism responsible for the USOC formation
[see Fig. \ref{FigID}(b)] has remained mainly unknown \cite{Buks_2951}. This
effect is partially attributed to the temperature dependence of the EDF
emission and absorption rates \cite{Franco_1090}. It has been shown that the
effective normalized EDF length, $\eta_{\mathrm{EDF}}$, which is defined as
$\eta_{\mathrm{EDF}}=l_{\mathrm{EDF}}/\left(  l_{\mathrm{E}}\log
\alpha_{\mathrm{FL}}\right)  $, where $l_{\mathrm{E}}$\ is the EDF emission
length, and $\alpha_{\mathrm{FL}}$ is the loss coefficient of the fiber loop,
becomes significantly smaller than unity in the low temperature region, in
which USOC is formed \cite{Buks_2951}. This transition from long to short EDF
(i.e., from $\eta_{\mathrm{EDF}}\gg1$ to $\eta_{\mathrm{EDF}}\ll1$) is
demonstrated by the open-loop measurements displayed in Figs. \ref{FigSetup}%
(c) and (d). In these measurements, a low-resolution optical spectrum analyzer
(OSA), connected to the left end of the EDF, is used to measured the EDF
transmittance. As seen from the plot in Fig. \ref{FigSetup} (c-d), the
open-loop transmittance near the pump wavelength of $980\operatorname{nm}$ is
significantly enhanced at low temperatures. The enhancement, which was
reported previously in Ref. \cite{Melle_2189}, is attributed to the increase
of the lifetime of excited states at low temperatures
\cite{Hehlen_9302,shafiei2022hole}, which, in turn, gives rise to a reduction
in the saturation value of the pump power.

The USOC $n$-th peak wavelength is denoted by $\lambda_{n}$, with
$n=0,1,2,\cdots$. The frequency associated with the $n$-th peak is
$f_{n}=c/\lambda_{n}$, where $c$ is the speed of light in vacuum, and the
corresponding normalized frequency detuning is $i_{n}\equiv\left(  f_{0}%
-f_{n}\right)  /f_{\mathrm{L}}$, where $f_{\mathrm{L}}=c/\left(
n_{\mathrm{F}}l_{\mathrm{L}}\right)  $ is the loop frequency, $n_{\mathrm{F}%
}=1.45$ is the fiber's refractive group index, and
$l_{\mathrm{L}}$ is the total length of the fiber loop. The detuning sequence
$\left\{  i_{n}\right\}  $ is found to be well fitted by an empirical law given by \cite{Buks_L051001}%
\begin{equation}
i_{n}=\nu\log p_{n}\ , \label{i_n}%
\end{equation}
where $\nu$ is a positive constant, and $p_{n}$ is the $n$-th prime number.
The degree of the agreement between the data and empiric law
(\ref{i_n}) has been evaluated in Ref. \cite{Buks_2951}.

\begin{figure}[ptb]
\begin{center}
\includegraphics[width=3.2in,keepaspectratio]{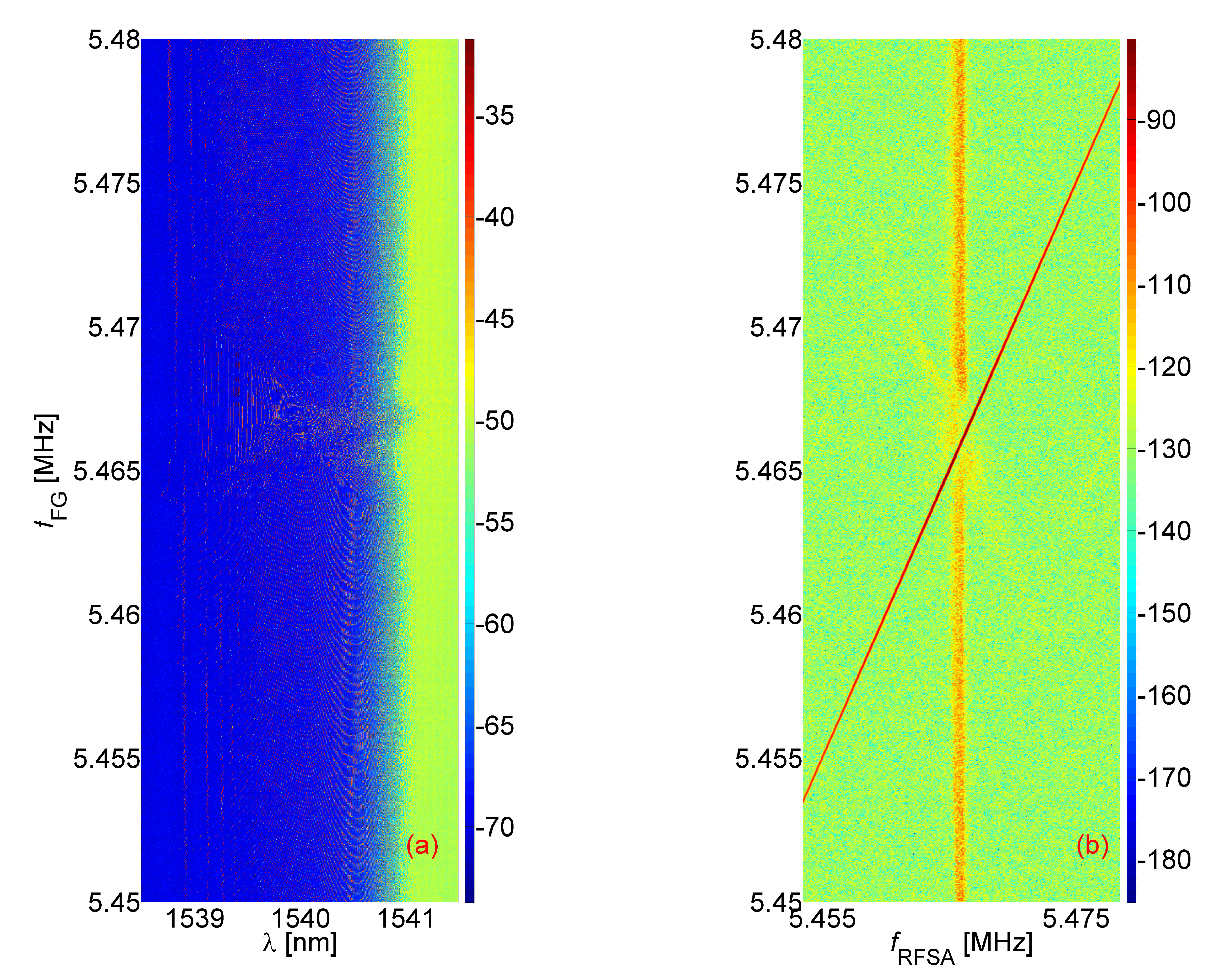}
\end{center}
\caption{EOM. (a) The optical spectrum measured by the POSA. (b) the RF
spectrum measured by the RFSA. The modulation frequency $f_{\mathrm{FG}}$ is
scanned near the loop frequency $f_{\mathrm{L}}=5.466 \operatorname{MHz}$. The
temperature is $2.8\operatorname{T}$, the diode current is $I_{\mathrm{D}%
}=120\operatorname{mA}$, and the EOM amplitude modulation depth is 0.5\%. In
both (a) and (b), the spectral power is presented in dBm.}%
\label{FigEOM}%
\end{figure}

USOC is highly sensitive to perturbations applied in the spectral band close
to the loop frequency $f_{\mathrm{L}}$. This is experimentally demonstrated by
installing an electro-optic modulator (EOM) into the fiber loop. The optical
and RF spectra are shown as a function of modulation frequency,
$f_{\mathrm{FG}}$, in plots (a) and (b), respectively, of Fig. \ref{FigEOM}
(for these measurements, the loop frequency is $f_{\mathrm{L}}=5.466
\operatorname{MHz}$). Note that EOM is installed in the fiber loop only for
the measurements shown in Fig. \ref{FigEOM}.

\textbf{Mode locking} -- The fiber loop under the study can be tuned into the
mode-locking regime. Current $I_{\mathrm{D}2}$ is applied for the measurements
shown in Figs. \ref{FigID}(c) and (d), which display a time trace (measured
with the use of the OS), and an optical spectrum, respectively, in the region
where the mode locking occurs (in these measurements, the OF is removed from
the experimental setup). The values $I_{\mathrm{D}1}$ and $I_{\mathrm{D}2}$
are denoted by the overlaid vertical dashed lines in Fig. \ref{FigID}(a). The
Kelly sidebands
\cite{Kelly_806,Gordon_91,Agrawal_Nonlinear_fiber_optics,Lin_045109} are
labeled in Fig. \ref{FigID}(d) by overlaid vertical dashed lines [see also
Fig. \ref{FigSetup}(b)]. While USOC is observed only at low temperatures, the
mode-locking is visible in the entire range from the room to base temperatures
[provided that a temperature-dependent adjustment is applied to both
polarization controllers PC1 and PC2].

The angular-frequency shift $\omega_{m}$ of the $m$-th Kelly sideband is%
\begin{equation}
\omega_{m}^{2}=2\pi\omega_{\mathrm{GDD}}^{2}m-t_{\mathrm{p}}^{-2}%
\ ,\label{Kelly sideband}%
\end{equation}
where $m$ is a positive integer, and the angular frequency $\omega
_{\mathrm{GDD}}$ is given by $\omega_{\mathrm{GDD}}^{-2}=\beta_{2,\mathrm{EDF}%
}l_{\mathrm{EDF}}+\beta_{2,\mathrm{SMF}}l_{\mathrm{SMF}}$. The group delay
dispersion (GDD) coefficient $\beta_{2}$ is $\beta_{2,\mathrm{SMF}%
}=18\operatorname{ps}^{2}\operatorname{km}^{-1}$ for the undoped
single-mode-fiber (SMF) section of the loop, and $\beta_{2,\mathrm{EDF}%
}=26\operatorname{ps}^{2}\operatorname{km}^{-1}$ for the EDF section,
$l_{\mathrm{EDF}}$ and $l_{\mathrm{SMF}}$ are, respectively, the EDF and SMF
lengths, and $t_{\mathrm{p}}$ is the pulse width. The overlaid vertical dashed
lines in Fig. \ref{FigSetup}(b) are calculated with the pulse width fitting
value of $t_{\mathrm{p}}=0.62\operatorname{ps}$.

Note that mode locking was not found after removing the PF from the fiber loop
(the search for mode locking was performed by varying the EDF temperature, the
diode current $I_{\mathrm{D}}$, and both PC1 and PC2 computer--controlled
settings). This observation suggests that the underlying mechanism responsible
for the mode locking in our setup is the nonlinear polarization rotation
\cite{Fermann_894}.

\textbf{The nonlinear Schr\"{o}dinger equation} -- Some
experimental results can be accounted for with the help of a simple model, in
which the light circulating in the fiber loop is represented by a scalar field
$\psi\left(  t,T\right)  $, which depends on the retarded time $t$ and the
time $T$ \cite{Agrawal_Nonlinear_fiber_optics,Kartashov_247}. The evolution is
modeled by the nonlinear Schr\"{o}dinger equation (NLSE)
\cite{Malomed_127802,Haus_1173} [see Eq. (\ref{SE |psi>}) in appendix
\ref{AppNSE}]:
\begin{equation}
i\frac{\mathrm{d}\psi}{\mathrm{d}T}=f_{\mathrm{L}}\left(  i\mathcal{G}%
+D\frac{\partial^{2}}{\partial t^{2}}+M\left\vert \psi\right\vert ^{2}\right)
\psi\;. \label{SE psi}%
\end{equation}
The net linear gain (the gain minus the loss) $\mathcal{G}$ is real, and the
complex coefficients $D=D_{2}+iD_{\mathrm{f}}$ and $M=M_{\delta}+iM_{\gamma}%
$\ characterize, respectively, the linear and cubic nonlinear response ($M$ is
proportional to the fiber's third-order susceptibility
\cite{Agrawal_Nonlinear_fiber_optics}), where $D_{2}=\beta_{2}/\left(
2f_{\mathrm{L}}\right)  $ is a normalized GDD coefficient, $D_{\mathrm{f}}$ is
the filter dispersion, $M_{\delta}$ is the self--phase modulation (SPM)
coefficient \cite{Montagna_871}, and $M_{\gamma}$ is the nonlinear gain
coefficient \cite{Jachpure_024007,Maeda_1187}. For $M_{\delta}<0$ ($M_{\delta
}>0$) self--focusing (self--defocusing) occurs. For saturable absorption (SA)
one has $M_{\gamma}>0$, whereas $M_{\gamma}<0$ occurs under the action of the
reverse saturable absorption (RSA). In the fiber-loop laser, SA commonly
promotes the mode locking and the formation of optical pulses
\cite{Kapitula_740,Dennis_1469,Huang_034009}, whereas effects which give rise
to RSA (e.g., two-photon absorption) suppress the mode locking
\cite{Wang_106805}. The changeover from SA to RSA has been observed in Ref.
\cite{Wei_7704}.

Modeling the mode-locking by dint of the NLSE (\ref{SE psi}) is reviewed in
appendices \ref{AppDS} and \ref{AppDisC}. Further, it is shown in appendices
\ref{AppME} and \ref{AppDampC} that the NLSE (\ref{SE psi}) can account for
the USOC formation, provided that some assumptions are applicable. In particular, the possibility to disregard both the SPM and dispersion (i.e.,
setting $M_{\delta}=0$ and $D=0$) in considered in appendix \ref{AppDampC}.
However, these assumptions are found to be inconsistent with our experimental
results (see appendix \ref{AppDampC}).

\textbf{The evolution of the pulse with the pump switched off} -- The role
played by the nonlinear response in establishing the mode-locking is explored
by modulating the diode current $I_{\mathrm{D}}$. For the plots shown in Fig. \ref{FigOffP}, the modulation abruptly switches
off the pump at time $\mathfrak{t}=0$. The subsequent decay was monitored
using three PDs. Figure \ref{FigOffP}(a) displays voltage $V_{\mathrm{P}}$
across a PD monitoring the optical power emitted from the
$980\operatorname{nm}$ diode. The fiber-loop signal decay is monitored by two
PDs. The signal measured by the faster one (having bandwidth of
$1\operatorname{GHz}$), which is capable to resolve individual optical pulses,
is shown in (b), whereas the plot in (c) is obtained using the slower one
(having bandwidth of $10\operatorname{kHz}$). The comparison between the plots
(b) and (c) reveals that the optical power (c) decays significantly faster
than the pulses' amplitude [which is visible in (b)]. This observation is
attributed to pulse narrowing which occurs in the course of the decay.

\begin{figure}[ptb]
\begin{center}
\includegraphics[width=3.2in,keepaspectratio]{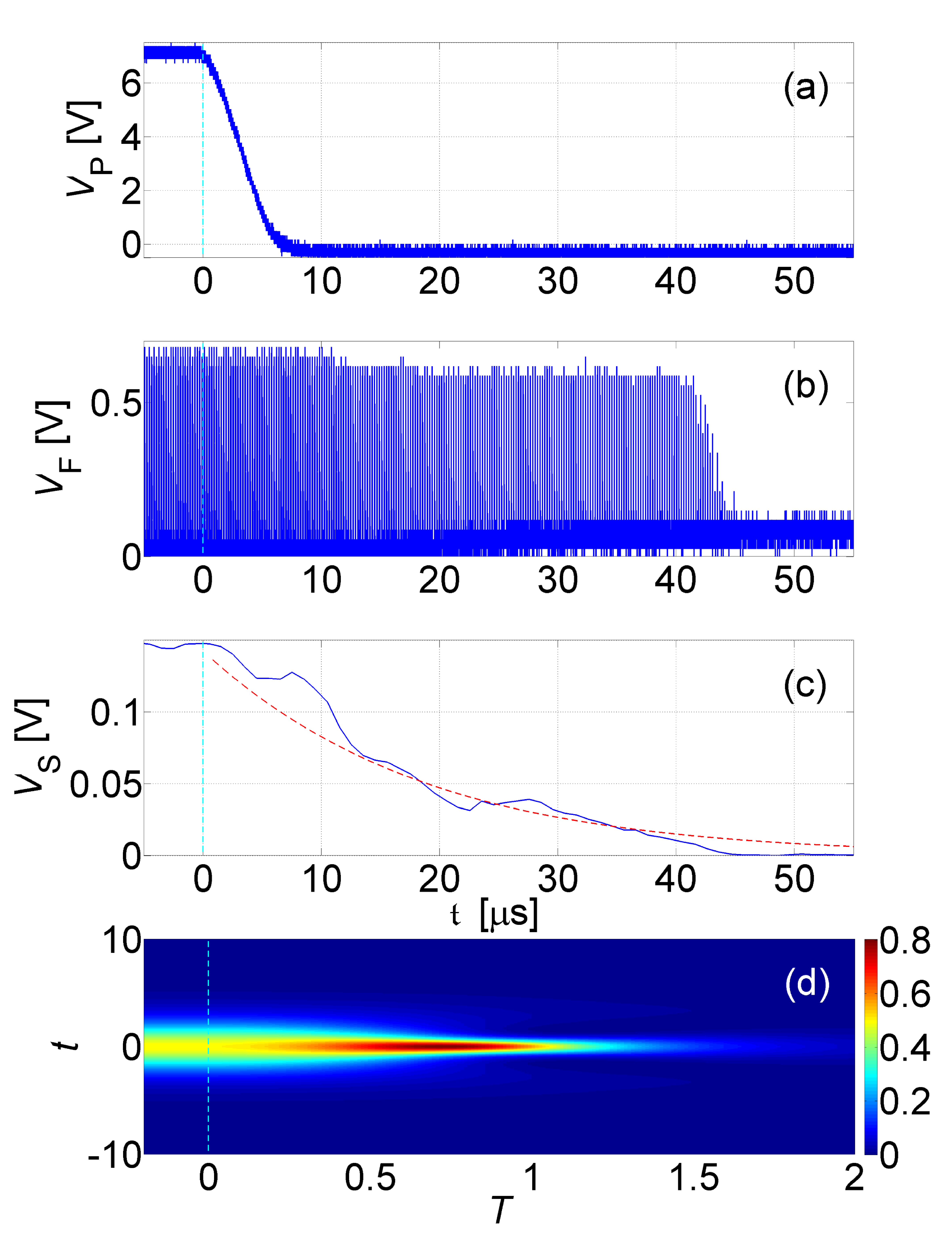}
\end{center}
\caption{{}The evolution of pulses with the pump switched off. Temperature is
$3.2\operatorname{K}$, and diode current prior to cutting off the pump at
$\mathfrak{t}=0$ is $I_{\mathrm{D}}=0.7 \operatorname{A} $. The pump power is
shown in (a), and the fiber-loop optical intensity in (b) and (c). The
frequency bandwidth of the photo detectors used in (a), (b) and (c), is $100
\operatorname{kHz} $, $1 \operatorname{GHz} $\ and $10 \operatorname{kHz} $,
respectively. The overlaid red dashed line in (c) represents the decay
function $I\left(  t\right)  $ defined in Eq. (\ref{I NLD}). Assumed parameters' values are $t_{\mathrm{d}}=17
\operatorname{\mu s}$ and $I_{\mathrm{s}}/I_{0}=10$. (d) The optical intensity
$\left\vert \psi\right\vert ^{2}$ is produced by the numerical solution of the
NLSE (\ref{SE psi}). Assumed parameters' values are
$W=2$, $1/f_{\mathrm{L}}=1$, $D_{2}=-1$, $D_{\mathrm{f}}=1$, $M_{\delta}=-1$
for $T<0$, $M_{\delta}=-10$ for $T>0$, $M_{\gamma}=1$, and $\mathcal{G}%
=-0.25$. The initial optical field $\psi$ is calculated using Eq.
(\ref{dissipative soliton}) of appendix \ref{AppDS}.}%
\label{FigOffP}%
\end{figure}

The overlaid red dashed line in Fig. \ref{FigOffP}(c) represents a fitting to
an empirical relation
\begin{equation}
I\left(  \mathfrak{t}\right)  =\frac{I_{\mathrm{s}}}{1+\frac{I_{\mathrm{s}%
}-I_{0}}{I_{0}}e^{\frac{\mathfrak{t}}{\mathfrak{t}_{\mathrm{d}}}}}\;.
\label{I NLD}%
\end{equation}
The decay function $I\left(  \mathfrak{t}\right)  $ given by Eq. (\ref{I NLD})
is derived from the logistic differential equation $\mathrm{d}I/\mathrm{d}%
\mathfrak{t}=-\mathfrak{t}_{\mathrm{d}}^{-1}\left(  1-I/I_{\mathrm{s}}\right)
I$, where $\mathfrak{t}_{\mathrm{d}}$ and $I_{\mathrm{s}}$ represent the
linear decay time and saturation intensity, respectively, and $I_{0}=I\left(
\mathfrak{t}=0\right)  $. The fit yields a positive value for the
parameter $I_{\mathrm{s}}$ (see the caption to Fig.
\ref{FigID}), which is consistent with the assumption that SA occurs in the
optical band where mode locking is observed.

For the case where both $D$ and $M$ are pure real, the narrowing effect can be
analytically modeled, provided that the adiabatic approximation is applicable
[see Eqs. (\ref{E2}) and (\ref{P2}) in appendix \ref{AppDisC}]. For the more
general case, NLSE (\ref{SE psi}) was numerically integrated. The color--coded
plot in Fig. \ref{FigOffP}(d) presents the numerically calculated optical
intensity $\left\vert \psi\right\vert ^{2}$ as a function of time $T$ and the
retarded time $t$. For this calculation, it is assumed that the SPM
coefficient $M_{\delta}$ abruptly changes at time $T=0$. Values of other
parameters are listed in the figure caption. The plot shown in Fig.
\ref{FigOffP}(d) demonstrates that an abrupt change in the nonlinear response
can give rise to narrowing, which mimics the experimental results shown in (b)
and (c) of Fig. \ref{FigOffP}. A possible mechanism, which may provide such an
abrupt change in the nonlinear response, is a $980\operatorname{nm}$
two--photon absorption occurring in the EDF \cite{Krug_1976,Xiao_1}. In
addition, the experimentally--observed narrowing occurring during the decay
process can be attributed to pumping--induced widening effect
\cite{Nakazawa_1075}.

\textbf{DOP} -- The POSA instrument, which is attached to the fiber loop [see
Fig. \ref{FigSetup}(a)], allows determining the polarization state of
individual USOC peaks \cite{Buks_486}. DOP that is extracted from the POSA
data is displayed in Fig. \ref{FigDOP}(b), for each of the USOC peaks shown in
Fig. \ref{FigDOP}(a). For this measurement, DOP values of about $0.9$ are
obtained [see Fig. \ref{FigDOP}(b)]. Much lower DOP values of about $0.12$ are
obtained when the PF is removed from the loop. As is demonstrated in the next
section, the relatively high DOP enables some useful applications.

\begin{figure}[ptb]
\begin{center}
\includegraphics[width=3.2in,keepaspectratio]{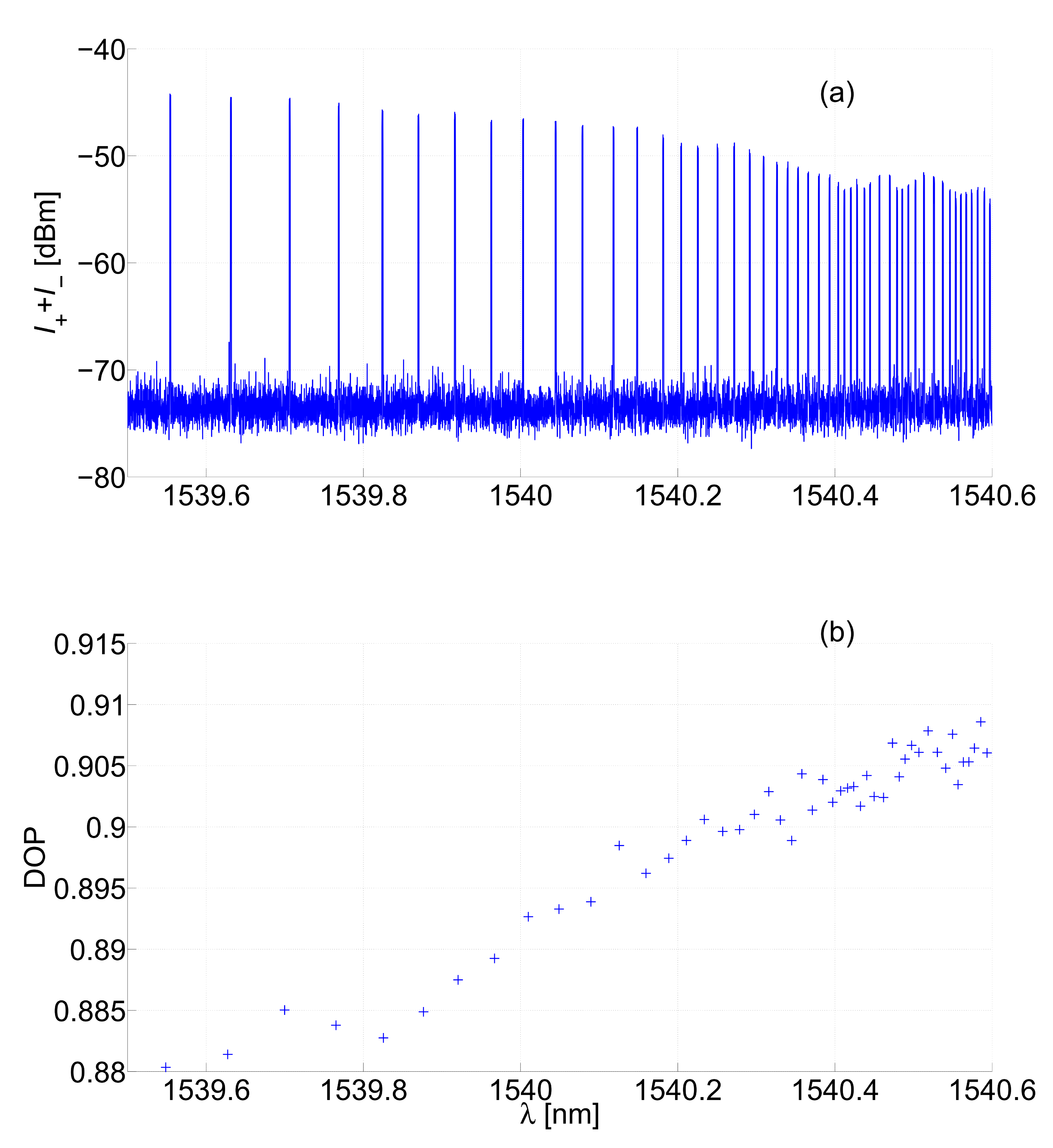}
\end{center}
\caption{{}DOP. The temperature is $3.5\operatorname{K}$, and the diode
current is $I_{\mathrm{D}}=0.3 \operatorname{A}$. The DOP is plotted in (b)
for all USOC peaks shown in (a).}%
\label{FigDOP}%
\end{figure}

\begin{figure}[ptb]
\begin{center}
\includegraphics[width=3.2in,keepaspectratio]{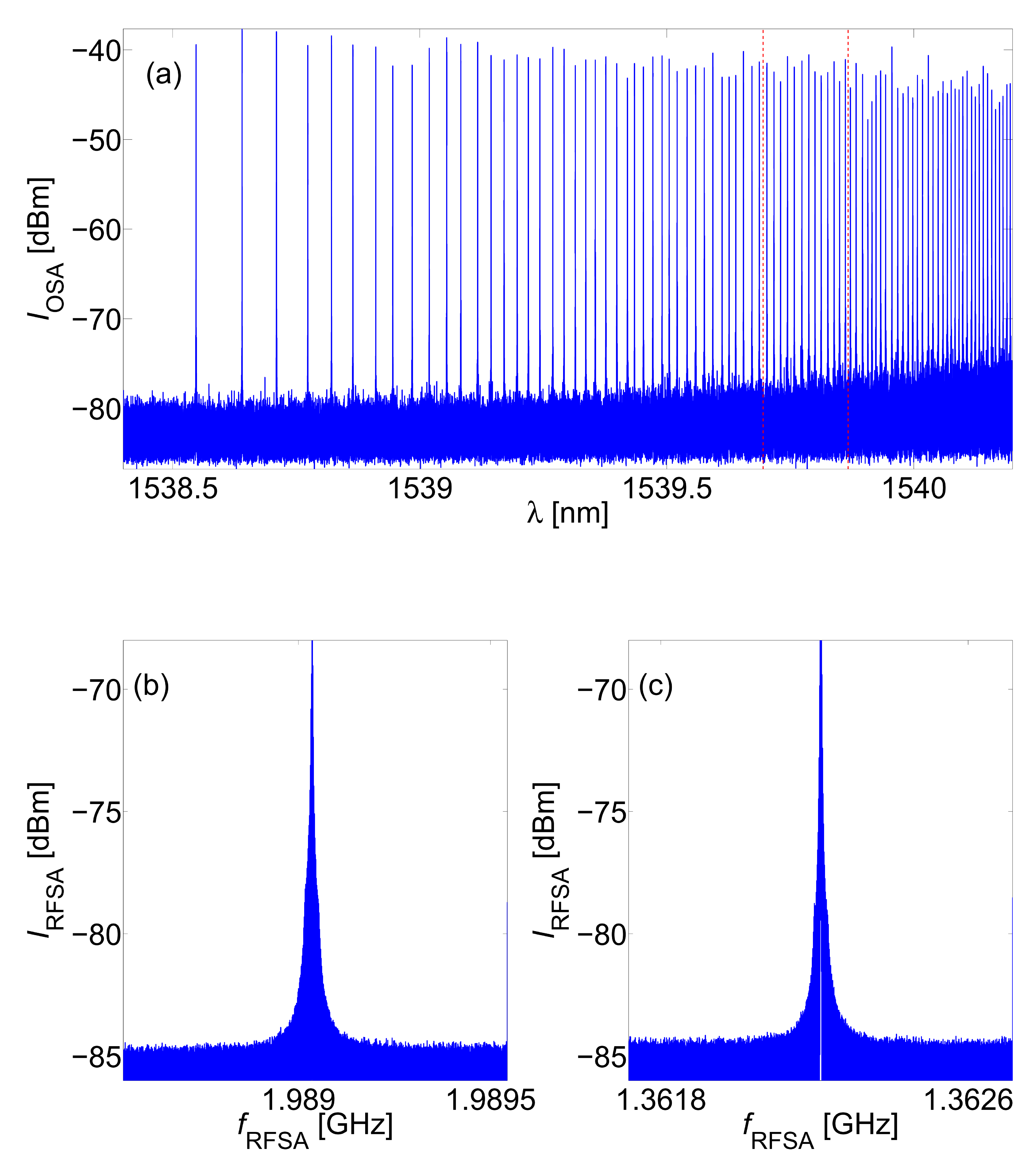}
\end{center}
\caption{{}RFSA. The temperature is $2.9\operatorname{K}$, and the diode
current is $I_{\mathrm{D}}=0.15\operatorname{A}$. The left and right overlaid
red vertical dashed lines in (a) label the OF wavelength $\lambda_{\mathrm{F}%
}$ used for the RFSA measurements shown in (b) and (c), respectively. In (b)
and (c), the averaging time is 25 minutes.}%
\label{FigRFSAOA}%
\end{figure}

\textbf{RF signal generation} -- Consider two USOC peaks having wavelengths
$\lambda$ and $\lambda+\delta_{\lambda}$, respectively. The beating frequency
$f_{\mathrm{b}}$ corresponding to this pair is $f_{\mathrm{b}}=\left(
c/\lambda^{2}\right)  \delta_{\lambda}+O\left(  \delta_{\lambda}^{2}\right)
$, where, in the optical band in which USOC occurs, i.e., for $\lambda
\simeq1540\operatorname{nm}$, $c/\lambda^{2}\simeq0.126\operatorname{GHz}%
\operatorname{pm} ^{-1}$. The plots shown in Fig. \ref{FigRFSAOA} demonstrate
an RFSA detection of USOC pairs' coherent beating. For these measurements the
OF is installed between the 10:90 OC and the PD [see Fig. \ref{FigSetup}(a)].
Note that, in addition, an optical amplifier [not shown in the sketch in Fig.
\ref{FigSetup}(a)] is added between the OF and the PD (to improve the
signal-to-noise ratio). The two overlaid vertical dashed red lines in Fig.
\ref{FigRFSAOA}(a) denote the OF central wavelength $\lambda_{\mathrm{F}}$ for
the plot shown in (b), for which $\lambda_{\mathrm{F}}=1539.7\operatorname{nm}%
$, and (c), for which $\lambda_{\mathrm{F}}=1539.9\operatorname{nm}$. The two
nearest-neighbors USOC peaks have the beating frequency $f_{\mathrm{b}%
}=1.989037\operatorname{GHz}$ and $f_{\mathrm{b}}=1.362216\operatorname{GHz}$,
for (b) and (c), respectively. The plots in (b) and (c) display the RFSA
spectral density $I_{\mathrm{RFSA}}$ averaged over 25 minutes. For both (b)
and (c), variation in the peak frequency is about $10\operatorname{kHz}$ over
this averaging period of 25 minutes. This frequency beating variation is
partially attributed to temperature fluctuations in the fiber-loop section
outside the cryostat.

\textbf{Summary} -- Further study is needed to reveal the underlying mechanism
responsible for the USOC formation. The device under the study allows a
fully--tunable multi--mode lasing regime with enhanced DOP and high stability.
These unique properties may open the way for novel USOC--based applications to
fields including spectroscopy, communications, and quantum data storage.

\textbf{Disclosures} -- The authors declare no conflicts of interest.

\appendix

\section{The nonlinear Schr\"{o}dinger equation (NLSE)}

\label{AppNSE}

The optical wave is represented by a state vector $\left\vert \psi\left(
t^{\prime},T\right)  \right\rangle $, which depends on the retarded time
$t^{\prime}$ and on full time $T$. The dynamics are governed by the NLSE,
\begin{equation}
i\frac{\mathrm{d}\left\vert \psi\right\rangle }{\mathrm{d}T}=H\left\vert
\psi\right\rangle \;, \label{SE |psi>}%
\end{equation}
where $H=\mathcal{H}+i\Theta$, and both $\mathcal{H}$ and $\Theta$ are
Hermitian, i.e. $\mathcal{H}^{\dag}=\mathcal{H}^{{}}$ and $\Theta^{\dag
}=\Theta^{{}}$. The state vector is represented by a wavefunction $\psi\left(
t^{\prime},T\right)  $ given by $\psi\left(  t^{\prime},T\right)
=\left\langle t^{\prime}\right.  \left\vert \psi\right\rangle $, where the
state $\left\vert t^{\prime}\right\rangle $ represents an optical wave
localized at the retarded time $t^{\prime}$, i.e., $\left\langle
t^{\prime\prime}\right.  \left\vert t^{\prime}\right\rangle =\delta\left(
t^{\prime}-t^{\prime\prime}\right)  $. The operators $\mathcal{H}$ and
$\Theta$ are $\mathcal{H}=f_{\mathrm{L}}\left(  -D_{2}p^{2}+M_{\delta
}\mathcal{N}\left(  \left\vert \psi\right\rangle \right)  \right)  $ and
$\Theta=f_{\mathrm{L}}\left(  \mathcal{G}-D_{\mathrm{f}}p^{2}+M_{\gamma
}\mathcal{N}\left(  \left\vert \psi\right\rangle \right)  \right)  $, where
$f_{\mathrm{L}}$ is the loop frequency, $D_{2}$ is group delay dispersion
(GDD), $M_{\delta}$ is the self--phase modulation (SPM), $\mathcal{G}$ is the
net gain (i.e., the gain minus loss), $D_{\mathrm{f}}$ is the filter
dispersion, and $M_{\gamma}$ is the nonlinear gain. The ``momentum" operator
$p$ satisfies the relation $\left\langle t^{\prime}\right\vert p\left\vert
\psi\right\rangle =-i\partial\psi/\partial t^{\prime}$, and the $\left\vert
\psi\right\rangle $--dependent operator $\mathcal{N}\left(  \left\vert
\psi\right\rangle \right)  $ is $\mathcal{N}\left(  \left\vert \psi
\right\rangle \right)  =\int\mathrm{d}t^{\prime\prime}\;\left\vert
\left\langle t^{\prime\prime}\right.  \left\vert \psi\right\rangle \right\vert
^{2}\left\vert t^{\prime\prime}\right\rangle \left\langle t^{\prime\prime
}\right\vert $. Multiplying the generalized Schr\"{o}dinger equation from the
left by $\left\langle t^{\prime}\right\vert $ yields Eq. (\ref{SE psi}). The
following relation holds, $f_{\mathrm{L}}^{-1}H=i\mathcal{G}-Dp^{2}%
+M\mathcal{N}\left(  \left\vert \psi\right\rangle \right)  $, where
$D=D_{2}+iD_{\mathrm{f}}$ and $M=M_{\delta}+iM_{\gamma}$. Note that the
Dirac's notation is employed above just for convenience, while the optical
wave is treated classically.

\section{Dissipative soliton}

\label{AppDS}

The NLSE (\ref{SE psi}) has a dissipative-soliton solution
\cite{Pereira_1733,Martinez_753,Haus_2068}
\begin{equation}
\psi=\frac{\sqrt{\frac{W}{2t_{\mathrm{p}}}}}{\cosh\frac{t}{t_{\mathrm{p}}}%
}\exp\left[  i\left(  \beta\log\left(  \frac{1}{\cosh\frac{t}{t_{\mathrm{p}}}%
}\right)  +\eta f_{\mathrm{L}}T\right)  \right]  \;,
\label{dissipative soliton}%
\end{equation}
where the chirp real coefficient $\beta$ is given by%
\begin{equation}
\beta=\frac{-3\chi\pm\sqrt{9\chi^{2}+8}}{2}\;, \label{be DS}%
\end{equation}
with
\begin{equation}
\frac{1}{\chi}=\frac{\frac{D_{2}}{D_{\mathrm{f}}}-\frac{M_{\delta}}{M_{\gamma
}}}{\frac{D_{2}}{D_{\mathrm{f}}}\frac{M_{\delta}}{M_{\gamma}}+1}\;,
\label{chi DS}%
\end{equation}
and the real variables $t_{\mathrm{p}}$ and $\eta$ satisfy the following
relations
\begin{align}
\frac{2\beta D_{\mathrm{f}}-\left(  1-\beta^{2}\right)  D_{2}}{t_{\mathrm{p}%
}^{2}}  &  =\eta\;,\\
\frac{-2\beta D_{2}-\left(  1-\beta^{2}\right)  D_{\mathrm{f}}}{t_{\mathrm{p}%
}^{2}}  &  =\mathcal{G}\;.
\end{align}
The energy corresponding to the dissipative solution is $W$, i.e.,
$\int_{-\infty}^{\infty}\mathrm{d}t\;\left\vert \psi\right\vert ^{2}=W$ [see
Eq. (\ref{dissipative soliton})]. Note that the chirp coefficient
vanishes, i.e. $\beta=0$, for both extreme cases $M_{\delta
}=D_{2}=0$ and $M_{\gamma}=D_{\mathrm{f}}=0$ [see Eqs. (\ref{be DS}) and
(\ref{chi DS})]. For the former case $t_{\mathrm{p}}=4D_{\mathrm{f}}/\left(
WM_{\gamma}\right)  $, and for the later one $t_{\mathrm{p}}=4D_{2}/\left(
WM_{\delta}\right)  $.

The complex NLSE (\ref{SE psi}) has non--diverging stable steady-state
solutions, provided that the damping is sufficiently strong. For the region
where the complex NLSE (\ref{SE psi}) becomes unstable, a finite and stable
steady-state solution can be obtained, provided that the gain-saturation
effect is taken into account. When the gain-relaxation time is long compared
to $f_{\mathrm{L}}^{-1}$, this effect is commonly accounted for by assuming
that $\mathcal{G}$ is a function of the total energy \cite{Jachpure_024007}.
Alternatively, the stability can be restored by implementing the fixed-energy
approximation. In this approximation, it is assumed that the total energy
$\int_{-\infty}^{\infty}\mathrm{d}t^{\prime}\;\left\vert \psi\left(
t^{\prime},T\right)  \right\vert ^{2}$ is $T$--independent. This approximation
can be implemented, by replacing NLSE (\ref{SE |psi>}) by a modified one
\cite{Buks_2400036}
\begin{equation}
i\frac{\mathrm{d}\left\vert \psi\right\rangle }{\mathrm{d}T}=\left(
\mathcal{H}+i\Theta-i\frac{\left\langle \psi\right\vert \Theta\left\vert
\psi\right\rangle }{\left\langle \psi\right.  \left\vert \psi\right\rangle
}\right)  \left\vert \psi\right\rangle \;. \label{MSE |psi>}%
\end{equation}
The term added to Eq. (\ref{MSE |psi>}) [compare with Eq. (\ref{SE |psi>})]
ensures that norm conservation condition $0=\left(  \mathrm{d}/\mathrm{d}%
T\right)  \left\langle \psi\right.  \left\vert \psi\right\rangle $ is satisfied.

\section{The case $D_{\mathrm{f}}=0$ and $M_{\gamma}=0$}

\label{AppDisC}

For the case where both $D$ and $M$ are pure real, the NLSE (\ref{SE psi})
becomes%
\begin{equation}
i\frac{\mathrm{d}\psi}{\mathrm{d}T}=f_{\mathrm{L}}\left(  i\mathcal{G}%
+D_{2}\frac{\partial^{2}}{\partial t^{2}}+M_{\delta}\left\vert \psi\right\vert
^{2}\right)  \psi\;. \label{SE psi D,M real}%
\end{equation}
We consider the case when $\mathcal{G}$, $D_{2}$ and $M_{\delta}$ are allowed
to be $T$ dependent. For this case, the total energy $W=\int_{-\infty}%
^{\infty}\mathrm{d}t\;\left\vert \psi\right\vert ^{2}$ evolves according to
[see Eq. (\ref{SE |psi>})]%
\begin{equation}
\frac{\mathrm{d}W}{\mathrm{d}T}=2f_{\mathrm{L}}\mathcal{G}\left(  T\right)  W.
\label{dW/dt}%
\end{equation}

Consider the case where $D_{2}$ and $M_{\delta}$ are both negative and slowly
varying functions of $T$. In the framework of the adiabatic perturbation
theory, the soliton solution is looked for as
\begin{equation}
\psi_{\mathrm{A}}\left(  t,T\right)  =\frac{A\left(  T\right)  \exp\left(
\frac{i}{2}\int_{0}^{T}\mathrm{d}T^{\prime}\;\upsilon^{2}\left(  T^{\prime
}\right)  \right)  }{\cosh\left(  \frac{\upsilon\left(  T\right)  t}%
{\sqrt{-2f_{\mathrm{L}}D_{2}\left(  T\right)  }}\right)  }\;, \label{sol2}%
\end{equation}
where $\upsilon\left(  T\right)  $ is real, and the soliton's amplitude is
\begin{equation}
A\left(  T\right)  =\frac{\upsilon\left(  T\right)  }{\sqrt{-f_{\mathrm{L}%
}M_{\delta}\left(  T\right)  }}\;. \label{A}%
\end{equation}
The energy $W$ of the approximate ansatz (\ref{sol2}) is%
\begin{equation}
W\left(  T\right)  =-\frac{2\sqrt{-\frac{2D_{2}\left(  T\right)
}{f_{\mathrm{L}}}}\upsilon\left(  T\right)  }{M_{\delta}\left(  T\right)  }\;.
\label{E2}%
\end{equation}
The approximate soliton solution (\ref{sol2}) is valid under the weak-gain
condition
\begin{equation}
\left\vert \mathcal{G}\left(  T\right)  \right\vert \ll\left\vert M_{\delta
}\left(  T\right)  \right\vert A^{2}\left(  T\right)  \;. \label{<<A^2}%
\end{equation}
In addition to this condition, the validity condition for the adiabatic
approximation is%
\begin{equation}
T_{0}\upsilon^{2}\left(  T\right)  \gg1\;, \label{adiabat}%
\end{equation}
where $T_{0}$ is a characteristic scale of the variation of the GDD and
nonlinearity coefficients (in the case of constant coefficients, $T_{0}
=\infty$).

In the adiabatic approximation, the evolution of $\upsilon\left(  T\right)  $
is governed by%
\begin{equation}
\frac{\mathrm{d\log}\upsilon}{\mathrm{d}T}=2f_{\mathrm{L}}\mathcal{G}\;.
\label{d nu / dz}%
\end{equation}
The soliton's peak power $P\left(  T\right)  \equiv-\upsilon^{2}\left(
T\right)  /\left(  f_{\mathrm{L}}M_{\delta}\left(  T\right)  \right)  $ is
obtained by the integration of Eq. (\ref{d nu / dz})%
\begin{equation}
\frac{P\left(  T\right)  }{P\left(  T=0\right)  }=\frac{M_{\delta}\left(
T=0\right)  }{M_{\delta}\left(  T\right)  }\exp\left(  4f_{\mathrm{L}}\int
_{0}^{T}\mathrm{d}T^{\prime}\;\mathcal{G}\left(  T^{\prime}\right)  \right)
\;. \label{P2}%
\end{equation}
Note that, in the adiabatic approximation, we have $W^{2}/P=8D_{2}\left(
T\right)  /M_{\delta}\left(  T\right)  $ [see Eq. (\ref{E2})].

\section{Mode expansion}

\label{AppME}

The Fourier expansion of $\psi\left(  t,T\right)  $ as%
\begin{equation}
\psi\left(  t,T\right)  =\sum_{m}a_{m}\left(  z\right)  e^{imx}\;,
\end{equation}
where $z=f_{\mathrm{L}}T$ and $x=\omega_{\mathrm{L}}t$, yields [see Eq.
(\ref{SE psi})]%
\begin{equation}
i\dot{a}_{m}=i\mathcal{G}a_{m}-D\omega_{\mathrm{L}}^{2}m^{2}a_{m}%
+M\sum_{m^{\prime}-m^{\prime\prime}+m^{\prime\prime\prime}=m^{{}}}%
a_{m^{\prime}}^{{}}a_{m^{\prime\prime}}^{\ast}a_{m^{\prime\prime\prime}}\;,
\label{a_m eom V1}%
\end{equation}
where $\omega_{\mathrm{L}}=2\pi f_{\mathrm{L}}$(the loop angular
frequency), and the overdot denotes a derivative with respect to
$z$.

To continue the analysis, we introduce the Wirtinger derivative $\partial
_{m}^{{}}$ and its conjugate $\partial_{m}^{\ast}$, which are defined by
$\partial_{m}^{{}}=\left(  1/2\right)  \left(  \partial/\partial\alpha
_{m}-i\partial/\partial\beta_{m}\right)  $ and $\partial_{m}^{\ast}=\left(
1/2\right)  \left(  \partial/\partial\alpha_{m}+i\partial/\partial\beta
_{m}\right)  $, respectively, where the complex $m$-th mode amplitude $a_{m}$
is expressed as $a_{m}=\alpha_{m}+i\beta_{m}$, and both $\alpha_{m}$ and
$\beta_{m}$ are real. The set of coupled equations of motion (\ref{a_m eom V1}%
) can be expressed in terms of the derivatives $\partial_{m}^{\ast}$ as
\cite{Ammari_29}%
\begin{equation}
\dot{a}_{m}=-\partial_{m}^{\ast}\mathbb{H}\;, \label{a_m eom V2}%
\end{equation}
where $\mathbb{H}=\mathbb{H}_{1}+i\mathbb{H}_{2}$, and the real $\mathbb{H}%
_{1}$ and imaginary $\mathbb{H}_{2}$ parts are given by%
\begin{align}
\mathbb{H}_{1}  &  =-\mathcal{G}Q_{1}+D_{\mathrm{f}}\omega_{\mathrm{L}}%
^{2}Q_{2}-\frac{M_{\gamma}}{2}Q_{3}\;,\\
\mathbb{H}_{2}  &  =-D_{2}\omega_{\mathrm{L}}^{2}Q_{2}+\frac{M_{\delta}}%
{2}Q_{3}\;,
\end{align}
where%
\begin{align}
Q_{1}  &  =\sum_{m^{\prime}}a_{m^{\prime}}^{\ast}a_{m^{\prime}}^{{}%
}\;,\label{Q1}\\
Q_{2}  &  =\sum_{m^{\prime}}m^{\prime2}a_{m^{\prime}}^{\ast}a_{m^{\prime}}%
^{{}}\;,\label{Q2}\\
Q_{3}  &  =\sum_{m^{\prime}-m^{\prime\prime}+m^{\prime\prime\prime}%
-m^{\prime\prime\prime\prime}=0}a_{m^{\prime}}^{{}}a_{m^{\prime\prime}}^{\ast
}a_{m^{\prime\prime\prime}}^{{}}a_{m^{\prime\prime\prime\prime}}^{\ast}\;.
\label{Q3}%
\end{align}

\section{Steady state}

\label{AppDampC}

Noise can be accounted for by replacing Eq. (\ref{a_m eom V2}) by%
\begin{equation}
\dot{a}_{m}=-\partial_{m}^{\ast}\mathbb{H}+\xi_{m}\;, \label{a_m eom V3}%
\end{equation}
where the added term $\xi_{m}$ on the right-hand side of Eq. (\ref{a_m eom V3}%
) represents random noise with a vanishing averaged value. We consider the
case with $D_{2}/D_{\mathrm{f}}=M_{\delta}/M_{\gamma}$. For that case, and in
the weak-noise limit, $\mathbb{H}_{1}$ is locally minimized in the steady
state [see, e.g., Eq. (7.258) in \cite{Buks_SPLN}]. Henceforth, it is further
assumed that both dispersion and SPM may be disregarded, i.e.
$D=0$ and $M_{\delta}=0$. These assumptions yield $\mathbb{H}%
_{{}}=\mathbb{H}_{1}=-\mathcal{G}Q_{1}-\left(  1/2\right)  M_{\gamma}Q_{3}$.

Note that the term $Q_{1}$ (\ref{Q1}) represents the total optical intensity.
In the fixed-intensity approximation it is assumed that $Q_{1}$ is a constant
(which is determined by the pump power). In this approximation, the term
$Q_{3}$ in the function $\mathbb{H}_{1}$\ can be replaced by $Q_{3}^{{}}%
-Q_{1}^{2}$ (note that constant terms in $\mathbb{H}$ do not affect the time
evolution). The following relation holds: $Q_{3}^{{}}=\sum_{\bar{n}\in
S_{\mathrm{c}}}a_{n^{\prime}}^{{}}a_{n^{\prime\prime}}^{\ast}a_{n^{\prime
\prime\prime}}^{{}}a_{n^{\prime\prime\prime\prime}}^{\ast}$ [see Eq.
(\ref{Q3})] and $Q_{1}^{2}=\sum_{\bar{n}\in S_{\mathrm{d}}}a_{n^{\prime}}^{{}%
}a_{n^{\prime\prime}}^{\ast}a_{n^{\prime\prime\prime}}^{{}}a_{n^{\prime
\prime\prime\prime}}^{\ast}$ [see Eq. (\ref{Q1})], where $\bar{n}=\left(
n^{\prime},n^{\prime\prime},n^{\prime\prime\prime},n^{\prime\prime\prime
}\right)  $ denotes a quadruple of integers, the set $S_{\mathrm{c}}%
\subset\mathbb{Z}^{4}$ is given by $S_{\mathrm{c}}=\left\{  \left(  n^{\prime
},n^{\prime\prime},n^{\prime\prime\prime},n^{\prime\prime\prime}\right)
\in\mathbb{Z}^{4}|n^{\prime}-n^{\prime\prime}+n^{\prime\prime\prime}%
-n^{\prime\prime\prime\prime}=0\right\}  $, and the set $S_{\mathrm{d}}%
\subset\mathbb{Z}^{4}$ is given by $S_{\mathrm{d}}=\left\{  \left(  n^{\prime
},n^{\prime\prime},n^{\prime\prime\prime},n^{\prime\prime\prime}\right)
\in\mathbb{Z}^{4}|\left(  n^{\prime},n^{\prime\prime}\right)  =\left(
n^{\prime\prime\prime\prime},n^{\prime\prime\prime}\right)  \right\}  $
($\mathbb{Z}$ denotes\ the set of all integers), and thus (the symbol
$\backslash$ denotes set subtraction)%
\begin{equation}
Q_{3}^{{}}-Q_{1}^{2}=\sum_{\bar{n}\in S_{\mathrm{c}}\backslash S_{\mathrm{d}}%
}a_{n^{\prime}}^{{}}a_{n^{\prime\prime}}^{\ast}a_{n^{\prime\prime\prime}}^{{}%
}a_{n^{\prime\prime\prime\prime}}^{\ast}\;. \label{Q3-Q_1^2}%
\end{equation}

For the evaluation of the term $Q_{3}^{{}}-Q_{1}^{2}$ given by Eq.
(\ref{Q3-Q_1^2}), it is useful to notice that the frequency spacings between
pairs of USOC peaks are all unique. Consider a quadruple of USOC peaks having
mode indices $m^{\prime}$, $m^{\prime\prime}$, $m^{\prime\prime\prime}$, and
$m^{\prime\prime\prime\prime}$. The empirical law given by Eq. (\ref{i_n})
yields%
\begin{equation}
i_{m^{\prime}}-i_{m^{\prime\prime}}+i_{m^{\prime\prime\prime}}-i_{m^{\prime
\prime\prime\prime}}=\nu\log\frac{p_{m^{\prime}}p_{m^{\prime\prime\prime}}%
}{p_{m^{\prime\prime}}p_{m^{\prime\prime\prime\prime}}}\;, \label{i_m USOC}%
\end{equation}
where $i_{m}=\left(  f_{0}-f_{m}\right)  /f_{\mathrm{L}}$ is the normalized
frequency detuning of the $m$-th mode. Note that $i_{m^{\prime}}%
-i_{m^{\prime\prime}}+i_{m^{\prime\prime\prime}}-i_{m^{\prime\prime
\prime\prime}}$ is proportional to $m^{\prime}-m^{\prime\prime}+m^{\prime
\prime\prime}-m^{\prime\prime\prime\prime}$, and thus the fundamental theorem
of arithmetics together with Eq. (\ref{i_m USOC}) imply that%
\begin{equation}
\bar{m}\equiv\left(  m^{\prime},m^{\prime\prime},m^{\prime\prime\prime
},m^{\prime\prime\prime}\right)  \notin S_{\mathrm{c}}\backslash
S_{\mathrm{d}}\;,
\end{equation}
provided that the pre--factor $\nu$ in Eq. (\ref{i_n}) is sufficiently large.
In other words, the term $Q_{3}^{{}}-Q_{1}^{2}$ [see Eq. (\ref{Q3-Q_1^2})]
vanishes when USOC is formed. Hence, the contribution of the intermode
coupling to $\mathbb{H}_{1}$ is minimized when USOC is formed, provided that
$M_{\gamma}<0$, i.e., RSA occurs in the USOC band (note that, generally,
$Q_{3}^{{}}-Q_{1}^{2}$ is non-negative). However, the applicability of the
above--discussed argument, which is based on the simplifying assumption that
the dispersion may be disregarded (in the band where USOC occurs), is
questionable, as it is not supported by dispersion measurements performed with
the system under the study in an open-loop configuration [see Fig. (6) in
\cite{Buks_2951}].

\bibliographystyle{ieeepes}
\bibliography{acompat,Eyal_Bib}

\newif\ifabfull\abfulltrue
\begin{thebibliography}{10}

\bibitem{Desurvire_246}
E~Desurvire, JL~Zyskind, and JR~Simpson,
\newblock ``Spectral gain holeburning at 1.53 m in erbium-doped fiber
  amplifiers'',
\newblock {\em IEEE Photon. Technol. Lett}, vol. 2, no. 4, pp. 246--248, 1990.

\bibitem{Nakazawa_613}
Masataka Nakazawa, Kazunori Suzuki, Hirokazu Kubota, and Yasuo Kimura,
\newblock ``Self-q-switching and mode locking in a 1.53-$\mu$m fiber ring laser
  with saturable absorption in erbium-doped fiber at 4.2 k'',
\newblock {\em Optics letters}, vol. 18, no. 8, pp. 613--615, 1993.

\bibitem{Thevenaz_22}
Luc Thevenaz, Alexandre Fellay, Massimo Facchini, Walter Scandale, Marc Nikles,
  and Philippe~A Robert,
\newblock ``Brillouin optical fiber sensor for cryogenic thermometry'',
\newblock in {\em Smart Structures and Materials 2002: Smart Sensor Technology
  and Measurement Systems}. International Society for Optics and Photonics,
  2002, vol. 4694, pp. 22--27.

\bibitem{Kobyakov_1}
Andrey Kobyakov, Michael Sauer, and Dipak Chowdhury,
\newblock ``Stimulated brillouin scattering in optical fibers'',
\newblock {\em Advances in optics and photonics}, vol. 2, no. 1, pp. 1--59,
  2010.

\bibitem{Le_3611}
Julien Le~Gou{\"e}t, J{\'e}r{\'e}my Oudin, Philippe Perrault, Alaeddine Abbes,
  Alice Odier, and Aliz{\'e}e Dubois,
\newblock ``On the effect of low temperatures on the maximum output power of a
  coherent erbium-doped fiber amplifier'',
\newblock {\em Journal of Lightwave Technology}, vol. 37, no. 14, pp.
  3611--3619, 2019.

\bibitem{Aubry_2100002}
Marine Aubry, Luciano Mescia, Adriana Morana, Thierry Robin, Arnaud Laurent,
  Julien Mekki, Emmanuel Marin, Youcef Ouerdane, Sylvain Girard, and Aziz
  Boukenter,
\newblock ``Temperature influence on the radiation responses of erbium-doped
  fiber amplifiers'',
\newblock {\em physica status solidi (a)}, vol. 218, no. 15, pp. 2100002, 2021.

\bibitem{Xi_1}
Qi~Xi, Shihai Wei, Chenzhi Yuan, Xueying Zhang, You Wang, Haizhi Song, Guangwei
  Deng, Bo~Jing, Daniel Oblak, and Qiang Zhou,
\newblock ``Experimental observation of coherent interaction between laser and
  erbium ions ensemble doped in fiber at sub 10 mk'',
\newblock {\em Science China Information Sciences}, vol. 63, pp. 1--7, 2020.

\bibitem{Wei_2209_00802}
Shi-Hai Wei, Bo~Jing, Xue-Ying Zhang, Jin-Yu Liao, Hao Li, Li-Xing You, Zhen
  Wang, You Wang, Guang-Wei Deng, Hai-Zhi Song, et~al.,
\newblock ``Storage of 1650 modes of single photons at telecom wavelength'',
\newblock {\em arXiv:2209.00802}, 2022.

\bibitem{Ortu_035024}
Antonio Ortu, Jelena~V Rakonjac, Adrian Holz{\"a}pfel, Alessandro Seri, Samuele
  Grandi, Margherita Mazzera, Hugues de~Riedmatten, and Mikael Afzelius,
\newblock ``Multimode capacity of atomic-frequency comb quantum memories'',
\newblock {\em Quantum Science and Technology}, vol. 7, no. 3, pp. 035024,
  2022.

\bibitem{Liu_2201_03692}
Duan-Cheng Liu, Pei-Yun Li, Tian-Xiang Zhu, Liang Zheng, Jian-Yin Huang,
  Zong-Quan Zhou, Chuan-Feng Li, and Guang-Can Guo,
\newblock ``On-demand storage of photonic qubits at telecom wavelengths'',
\newblock {\em arXiv:2201.03692}, 2022.

\bibitem{Bornadel_2412_16013}
Mahdi Bornadel, Sara~Shafiei Alavijeh, Farhad Rasekh, Nasser~Gohari Kamel,
  Faezeh~Kimiaee Asadi, Erhan Saglamyurek, Daniel Oblak, and Christoph Simon,
\newblock ``Hole burning experiments and modeling in erbium-doped silica glass
  fibers down to millikelvin temperatures: evidence for ultra-long population
  storage'',
\newblock {\em arXiv:2412.16013}, 2024.

\bibitem{Tittel_2501_06110}
Wolfgang Tittel, Mikael Afzelius, Adam Kinos, Lars Rippe, and Andreas Walther,
\newblock ``Quantum networks using rare-earth ions'',
\newblock {\em arXiv:2501.06110}, 2025.

\bibitem{Jing_031304}
Bo~Jing, Shihai Wei, Longyao Zhang, Dianli Zhou, Yuxing He, Xihua Zou, Wei Pan,
  Hai-Zhi Song, and Lianshan Yan,
\newblock ``Approaching scalable quantum memory with integrated atomic
  devices'',
\newblock {\em Applied Physics Reviews}, vol. 11, no. 3, pp. 031304, 2024.

\bibitem{Gupta_044029}
Shobhit Gupta, Xuntao Wu, Haitao Zhang, Jun Yang, and Tian Zhong,
\newblock ``Robust millisecond coherence times of erbium electron spins'',
\newblock {\em Physical Review Applied}, vol. 19, no. 4, pp. 044029, 2023.

\bibitem{Saglamyurek_241111}
Erhan Saglamyurek, Thomas Lutz, Lucile Veissier, Morgan~P Hedges, Charles~W
  Thiel, Rufus~L Cone, and Wolfgang Tittel,
\newblock ``Efficient and long-lived zeeman-sublevel atomic population storage
  in an erbium-doped glass fiber'',
\newblock {\em Physical Review B}, vol. 92, no. 24, pp. 241111, 2015.

\bibitem{Shafiei_F2A}
Sara Shafiei, Erhan Saglamyurek, and Daniel Oblak,
\newblock ``Hour-long decay-time of erbium spins in an optical fiber at
  milli-kelvin temperatures'',
\newblock in {\em Quantum Information and Measurement}. Optical Society of
  America, 2021, pp. F2A--4.

\bibitem{Haken_laser}
Hermann Haken,
\newblock {\em Laser light dynamics}, vol.~1,
\newblock North-Holland Amsterdam, 1985.

\bibitem{Semenov_23}
Andriy~O Semenov, Serhii Tsyrulnyk, Olena~O Semenova, Serhii Baraban, and Anton
  Khloba,
\newblock ``Dynamic random access memory based on fiber optic lines for optical
  computers. computer modeling.'',
\newblock in {\em COLINS (2)}, 2024, pp. 23--34.

\bibitem{Yamashita_1298}
S~Yamashita and K~Hotate,
\newblock ``Multiwavelength erbium-doped fibre laser using intracavity etalon
  and cooled by liquid nitrogen'',
\newblock {\em Electronics Letters}, vol. 32, no. 14, pp. 1298--1299, 1996.

\bibitem{Liu_102988}
Haochong Liu, Wei He, Yantao Liu, Yunhui Dong, and Lianqing Zhu,
\newblock ``Erbium-doped fiber laser based on femtosecond laser inscribed fbg
  through fiber coating for strain sensing in liquid nitrogen environment'',
\newblock {\em Optical Fiber Technology}, vol. 72, pp. 102988, 2022.

\bibitem{Lopez_085401}
J~Lopez, H~Kerbertt, M~Plata, E~Hernandez, and S~Stepanov,
\newblock ``Two-wave mixing in erbium-doped-fibers with spectral-hole burning
  at 77k'',
\newblock {\em Journal of Optics}, vol. 22, no. 8, pp. 085401, 2020.

\bibitem{Buks_L051001}
Eyal Buks,
\newblock ``Tunable multimode lasing in a fiber ring'',
\newblock {\em Physical Review Applied}, vol. 19, no. 5, pp. L051001, 2023.

\bibitem{Hofer_720}
Martin Hofer, MH~Ober, F~Haberl, and ME~Fermann,
\newblock ``Characterization of ultrashort pulse formation in passively
  mode-locked fiber lasers'',
\newblock {\em IEEE journal of quantum electronics}, vol. 28, no. 3, pp.
  720--728, 1992.

\bibitem{Fermann_894}
ME~Fermann, MJ~Andrejco, Y~Silberberg, and ML~Stock,
\newblock ``Passive mode locking by using nonlinear polarization evolution in a
  polarization-maintaining erbium-doped fiber'',
\newblock {\em Optics letters}, vol. 18, no. 11, pp. 894--896, 1993.

\bibitem{Buks_486}
Eyal Buks,
\newblock ``Polarimeter optical spectrum analyzer'',
\newblock {\em Photonics}, vol. 11, no. 6, pp. 486, 2024.

\bibitem{Buks_2951}
Eyal Buks,
\newblock ``Intermode coupling in a fiber loop laser at low temperatures'',
\newblock {\em Journal of Lightwave Technology}, vol. 42, pp. 2951, 2024.

\bibitem{Franco_1090}
Pierluigi Franco, Michele Midrio, A~Tozzato, Marco Romagnoli, and F~Fontana,
\newblock ``Characterization and optimization criteria for filterless
  erbium-doped fiber lasers'',
\newblock {\em JOSA B}, vol. 11, no. 6, pp. 1090--1097, 1994.

\bibitem{Melle_2189}
Sonia Melle, Oscar~G Calder{\'o}n, Miguel~A Ant{\'o}n, Fernando Carreno, and
  Ana Egatz-G{\'o}mez,
\newblock ``Spectral hole burning in erbium-doped fibers for slow light'',
\newblock {\em JOSA B}, vol. 29, no. 8, pp. 2189--2198, 2012.

\bibitem{Hehlen_9302}
Markus~P Hehlen, Nigel~J Cockroft, TR~Gosnell, and Allan~J Bruce,
\newblock ``Spectroscopic properties of er 3+-and yb 3+-doped soda-lime
  silicate and aluminosilicate glasses'',
\newblock {\em Physical Review B}, vol. 56, no. 15, pp. 9302, 1997.

\bibitem{shafiei2022hole}
Sara Shafiei~Alavijeh,
\newblock ``Hole burning spectroscopy of erbium-doped optical fibre for
  applications in quantum networks'',
\newblock 2022.

\bibitem{Kelly_806}
SMJ Kelly,
\newblock ``Characteristic sideband instability of periodically amplified
  average soliton'',
\newblock {\em Electronics Letters}, vol. 28, no. 8, pp. 806--807, 1992.

\bibitem{Gordon_91}
James~P Gordon,
\newblock ``Dispersive perturbations of solitons of the nonlinear
  schr{\"o}dinger equation'',
\newblock {\em JOSA B}, vol. 9, no. 1, pp. 91--97, 1992.

\bibitem{Agrawal_Nonlinear_fiber_optics}
Govind~P Agrawal,
\newblock ``Nonlinear fiber optics'',
\newblock in {\em Nonlinear Science at the Dawn of the 21st Century}, pp.
  195--211. Springer, 2000.

\bibitem{Lin_045109}
Yung-Hsiang Lin and Gong-Ru Lin,
\newblock ``Kelly sideband variation and self four-wave-mixing in femtosecond
  fiber soliton laser mode-locked by multiple exfoliated graphite
  nano-particles'',
\newblock {\em Laser Physics Letters}, vol. 10, no. 4, pp. 045109, 2013.

\bibitem{Kartashov_247}
Yaroslav~V Kartashov, Boris~A Malomed, and Lluis Torner,
\newblock ``Solitons in nonlinear lattices'',
\newblock {\em Reviews of Modern Physics}, vol. 83, no. 1, pp. 247--305, 2011.

\bibitem{Malomed_127802}
Boris~A Malomed,
\newblock ``New findings for the old problem: Exact solutions for domain walls
  in coupled real ginzburg-landau equations'',
\newblock {\em Physics Letters A}, vol. 422, pp. 127802, 2022.

\bibitem{Haus_1173}
Herman~A Haus,
\newblock ``Mode-locking of lasers'',
\newblock {\em IEEE Journal of Selected Topics in Quantum Electronics}, vol. 6,
  no. 6, pp. 1173--1185, 2000.

\bibitem{Montagna_871}
M~Montagna, S~Selleri, and M~Zoboli,
\newblock ``Nonlinear refractive index in erbium-doped optical amplifiers'',
\newblock {\em Optical and quantum electronics}, vol. 27, pp. 871--880, 1995.

\bibitem{Jachpure_024007}
Deeksha Jachpure and R~Vijaya,
\newblock ``Saturable absorption and its consequent effects in bistable
  erbium-doped fiber ring laser'',
\newblock {\em Journal of Optics}, vol. 24, no. 2, pp. 024007, 2022.

\bibitem{Maeda_1187}
Yoshinobu Maeda,
\newblock ``Mechanism of the negative nonlinear absorption effect in a
  five-level system of the er 3+ ion'',
\newblock {\em Journal of applied physics}, vol. 83, no. 3, pp. 1187--1194,
  1998.

\bibitem{Kapitula_740}
Todd Kapitula, J~Nathan Kutz, and Bj{\"o}rn Sandstede,
\newblock ``Stability of pulses in the master mode-locking equation'',
\newblock {\em JOSA B}, vol. 19, no. 4, pp. 740--746, 2002.

\bibitem{Dennis_1469}
Michael~L Dennis and Irl~N Duling,
\newblock ``Experimental study of sideband generation in femtosecond fiber
  lasers'',
\newblock {\em IEEE Journal of Quantum electronics}, vol. 30, no. 6, pp.
  1469--1477, 1994.

\bibitem{Huang_034009}
Shen Huang, Guodong Shao, Yufeng Song, Luming Zhao, Deyuan Shen, and Dingyuan
  Tang,
\newblock ``Dark solitons embedded in a stable periodic pulse train emitted by
  a fiber ring laser'',
\newblock {\em Journal of Physics: Photonics}, vol. 2, no. 3, pp. 034009, 2020.

\bibitem{Wang_106805}
Gang Wang, Yuxuan Ma, Ce~Shang, Haojing Huang, Zherui Lu, Shuaixin Wang,
  Jingxuan Sun, Chenghong Zhang, and Bo~Fu,
\newblock ``Influence of reverse saturable absorption effect on conventional
  and dissipative solitons fiber lasers'',
\newblock {\em Optics \& Laser Technology}, vol. 137, pp. 106805, 2021.

\bibitem{Wei_7704}
Rongfei Wei, Hang Zhang, Xiangling Tian, Tian Qiao, Zhongliang Hu, Zhi Chen,
  Xin He, Yongze Yu, and Jianrong Qiu,
\newblock ``Mos 2 nanoflowers as high performance saturable absorbers for an
  all-fiber passively q-switched erbium-doped fiber laser'',
\newblock {\em Nanoscale}, vol. 8, no. 14, pp. 7704--7710, 2016.

\bibitem{Krug_1976}
Peter~A Krug, Mark~G Sceats, GR~Atkins, SC~Guy, and Simon~B Poole,
\newblock ``Intermediate excited-state absorption in erbium-doped fiber
  strongly pumped at 980 nm'',
\newblock {\em Optics letters}, vol. 16, no. 24, pp. 1976--1978, 1991.

\bibitem{Xiao_1}
Gui Xiao, Ghazal Fallah~Tafti, Amirhassan Zareanborji, Anahita Ghaznavi, and
  Qiancheng Zhao,
\newblock ``Measurement of active optical fibers'',
\newblock in {\em Handbook of Optical Fibers}, pp. 1--38. Springer, 2019.

\bibitem{Nakazawa_1075}
Masataka Nakazawa, Hirokazu Kubota, Akio Sahara, and Kohichi Tamura,
\newblock ``Time-domain abcd matrix formalism for laser mode-locking and
  optical pulse transmission'',
\newblock {\em IEEE Journal of Quantum Electronics}, vol. 34, no. 7, pp.
  1075--1081, 1998.

\bibitem{Pereira_1733}
N.R. Pereira and Lennart Stenflo,
\newblock ``Nonliner schrodinger equation including growth and damping'',
\newblock {\em The Physics of Fluids}, vol. 20, pp. 1733, 1977.

\bibitem{Martinez_753}
Oscar~E Mart{\'\i}nez, Richard~L Fork, and James~P Gordon,
\newblock ``Theory of passively mode-locked lasers for the case of a nonlinear
  complex-propagation coefficient'',
\newblock {\em Journal of the Optical Society of America B}, vol. 2, no. 5, pp.
  753--760, 1985.

\bibitem{Haus_2068}
Hermann~A Haus, James~G Fujimoto, and Erich~P Ippen,
\newblock ``Structures for additive pulse mode locking'',
\newblock {\em Journal of the Optical Society of America B}, vol. 8, no. 10,
  pp. 2068--2076, 1991.

\bibitem{Buks_2400036}
Eyal Buks,
\newblock ``Spontaneous disentanglement and thermalization'',
\newblock {\em Advanced Quantum Technologies}, p. 2400036, 2024.

\bibitem{Ammari_29}
Zied Ammari and Vedran Sohinger,
\newblock ``Gibbs measures as unique kms equilibrium states of nonlinear
  hamiltonian pdes'',
\newblock {\em Revista Matem{\'a}tica Iberoamericana}, vol. 39, no. 1, pp.
  29--90, 2022.

\bibitem{Buks_SPLN}
Eyal Buks,
\newblock {\em Statistical physics - Lecture Notes},
\newblock http://buks.net.technion.ac.il/teaching/, 2025.

\end{thebibliography}

\end{document}